\documentclass[nofootinbib,preprint,11pt]{revtex4-1}

\usepackage{tikz}
\usepackage{subfigure}
\usepackage[section]{placeins}
\usepackage{fancyhdr}
\usepackage{array}
\usepackage{mathtools}
\usepackage{amssymb}
\usepackage{graphicx}
\usepackage{epsfig}
\usepackage{wrapfig}
\usepackage{graphics}
\usepackage{slashed}	
\usepackage{enumitem}
\usepackage{color}
\definecolor{SeaBlue}{rgb}{0.1,0.4,0.85}
\definecolor{DarkBlue}{rgb}{0.1,0.3,0.65}
\definecolor{Maroon}{rgb}{0.6,0.2,0.2}
\definecolor{SeaGreen}{rgb}{0.2,0.4,0.2}
\definecolor{Purple}{rgb}{0.7,0.1,0.6}
\definecolor{Red}{rgb}{0.8,0.2,0.2}
\definecolor{Black}{rgb}{0.0,0.0,0.0}
\definecolor{ZQS}{rgb}{0.01,0.227,0.451}
\definecolor{TS}{rgb}{0.588,0.31,0.05}
\usepackage[colorlinks]{hyperref}
\hypersetup{
     colorlinks   = true,
     citecolor    = red,
	linkcolor=SeaBlue
}
\usepackage[T1]{fontenc}
\usepackage{titlesec}
\titleformat{\section}
  { \color{ZQS} \normalfont \scshape}
  {\thesection}{1em}{\bf \MakeUppercase}
\titleformat{\subsection} { \color{SeaBlue}  \normalfont\scshape}{\thesubsection}{1em}{\bf }{}
\titleformat{\subsubsection}
  { \color{SeaBlue}  \normalfont\itshape}{\thesubsubsection }{1em}{}{}

\usetikzlibrary{decorations.pathmorphing}	
\usetikzlibrary{decorations.markings}
\tikzset{
    vector/.style={decorate, decoration={snake}, draw},
	provector/.style={decorate, decoration={snake,amplitude=2.5pt}, draw},
	antivector/.style={decorate, decoration={snake,amplitude=-2.5pt}, draw},
    fermion/.style={draw=black, postaction={decorate},
        decoration={markings,mark=at position .55 with {\arrow[draw=black]{>}}}},
    fermionbar/.style={draw=black, postaction={decorate},
        decoration={markings,mark=at position .55 with {\arrow[draw=black]{<}}}},
    fermionnoarrow/.style={draw=black},
    gluon/.style={decorate, draw=black,
        decoration={coil,amplitude=4pt, segment length=5pt}},
    scalar/.style={dashed,draw=black, postaction={decorate},
        decoration={markings,mark=at position .55 with {\arrow[draw=black]{>}}}},
    scalarbar/.style={dashed,draw=black, postaction={decorate},
        decoration={markings,mark=at position .55 with {\arrow[draw=black]{<}}}},
    scalarnoarrow/.style={dashed,draw=black},
    electron/.style={draw=black, postaction={decorate},
        decoration={markings,mark=at position .55 with {\arrow[draw=black]{>}}}},
	bigvector/.style={decorate, decoration={snake,amplitude=4pt}, draw},
}

\tikzstyle{block} = [draw, rectangle, 
    minimum height=3em, minimum width=6em]

\definecolor{cerulean}{rgb}{0., 0.52,0.65}

\newcolumntype{P}[1]{>{\raggedright\arraybackslash}p{#1}}

\usepackage{aas_macros}

\makeatletter
\def\l@subsubsection#1#2{}
\makeatother

\begin{document}

\title{\Large Leptophilic Composite Asymmetric Dark Matter and its Detection}
\author{\large Mengchao Zhang}
\email{mczhang@jnu.edu.cn}
\affiliation{\large Department of Physics and Siyuan Laboratory, Jinan University, Guangzhou 510632, P.R. China }
\date{\today}
\begin{abstract} 
\normalsize We propose a model which explains the baryon asymmetry of the universe and dark matter relic density at the same time. In this model, dark matter candidate is the dark baryon composed by dark quarks. A scalar mediator, which couples to the standard model leptons and dark quarks, is introduced to generate the asymmetry of baryon and dark baryon simultaneously. Direct detection and collider detection of this model are studied. We find that current underground direct detection experiments and LHC can hardly detect this model. But future lepton colliders, such as CEPC, have great potential to detect a large portion of the model parameter space by ``displaced lepton jet'' signal. 
\end{abstract}
\maketitle

\newpage

\section{Introduction \label{sec:introduction} }

The observed baryon asymmetry of the universe (BAU) is a main puzzle in modern cosmology and particle physics. 
The ratio of the baryon number density to entropy density, $Y_{\Delta B} = (n_B - n_{\bar{B}})/s \approx 8\times 10^{-11}$, is measured by the cosmic microwave backgound (CMB)~\cite{Akrami:2018vks} and Big-Bang nucleosynthesis (BBN)~\cite{Fields:2019pfx}.
To explain BAU, three Sakharov conditions~\cite{Sakharov:1967dj} need to be satisfied: baryon number violation, C \& CP violation, and departure from thermal equilibrium. 
Successful baryogenesis mechanisms include electroweak baryogenesis~\cite{Kuzmin:1985mm,Shaposhnikov:1986jp,Shaposhnikov:1987tw}, leptogenesis~\cite{Fukugita:1986hr}, or the Affleck-Dine mechanism~\cite{Affleck:1984fy,Dine:1995kz}. 
Recent review see Ref.~\cite{Bodeker:2020ghk}. 

Another important issue in modern cosmology is the nature of dark matter (DM). 
So far, we only know the existence of DM through its gravitational effects~\cite{Ade:2015xua}.  
The lack of observations other than the gravitational effects makes the DM model building highly speculative.  
Weakly interacting massive particle (WIMP) DM, which can naturally explain observed DM relic density by freeze-out mechanism, has been intensively searched for by indirect detection (ID) experiments, direct detection (DD) experiments, and collider experiments~\cite{DirectSearch1,DirectSearch2,LHCSearch1,LHCSearch2,LHCSearch3,LHCSearch4,LHCSearch5}. 
But definite observation evidence of WIMP has not been seen in above experiments. 

Motivated by the null results in WIMP DM searches and the coincidence that the abundance of baryon and DM are very close ($\Omega_{\text{DM}} \simeq 5\Omega_\text{B}$), it is natural to assume that the origin of baryon and DM abundance might be related to each other. 
To be more specific, the DM abundance is determined by the asymmetry between DM and anti-DM, and the asymmetry in dark sector and visible sector are from the same source. 
Such an asymmetric dark matter (ADM) paradigm is very different from WIMP in terms of model building and phenomenology study~\cite{Nussinov:1985xr,Kaplan:1991ah,Barr:1990ca,Barr:1991qn,Dodelson:1991iv,Fujii:2002aj,Kitano:2004sv,Farrar:2005zd,Kitano:2008tk,Gudnason:2006ug,Kaplan:2009ag,Shelton:2010ta,Davoudiasl:2010am,Huang:2017kzu,Buckley:2010ui,Cohen:2010kn,Frandsen:2011kt,Petraki:2013wwa,Zurek:2013wia,Ibe:2018juk,Ibe:2018tex,An:2009vq,Falkowski:2011xh,Bai:2013xga,Alves:2009nf,Alves:2010dd,Blennow:2012de}. 

In this paper we focus on the composite ADM model~\cite{Ibe:2018juk,Ibe:2018tex,Alves:2009nf,Alves:2010dd,Beylin:2020bsz,Khlopov:1989fj,Blinnikov:1983gh,Blinnikov:1982eh}. 
In composite ADM model, there is a confined QCD-like strong interaction in the dark sector, and the DM candidate is the lightest dark baryon. 
Thus the mass of DM is basically determined by the dark confinement scale $\Lambda'_{\text{QCD}}$~\footnote{In this paper we use $'$ to denote objects in the dark sector.}. 
Furthermore, we use the decay of mediator particle (labeled as $\Phi$) to generate the asymmetry in both visible and dark sector, as proposed in~\cite{Bai:2013xga}. 
The mediator $\Phi$ carries particle numbers in both visible and dark sector. 
Introducing such a ``bi-charged'' mediator have many advantages. 
Firstly, the asymmetry in both visible and dark sector comes from the asymmetry of mediator $\Phi$, and thus the particle number densities in two sectors are automatically connected to each other. 
Secondly, this mediator provides a portal which release the entropy in the dark sector to visible sector, and thus prevent the dark sector from being hot~\cite{Blennow:2012de,Ibe:2018juk}. 
Finally, the mass of mediator $\Phi$ can be quite low, which makes this model detectable at current or coming experiments. 

In previous work~\cite{Bai:2013xga}, mediator $\Phi$ is chosen to couple to a dark quark (labeled as $q'$) and a SM quark. 
Being charged under the SM QCD, $\Phi$ can be copiously produced at the LHC, and the dark quark $q'$ in the final state eventually evolves to a so-called ``dark-jet''. 
Such a novel collider signature has induced many studies in recent years~\cite{Strassler:2006im,Strassler:2006ri,Strassler:2006qa,Han:2007ae,Verducci:2011zz,Chan:2011aa,Strassler:2008fv,Cohen:2015toa,Bai:2013xga,Schwaller:2015gea,Beauchesne:2017yhh,Pierce:2017taw,Beauchesne:2018myj,Renner:2018fhh,Park:2017rfb,Mies:2020mzw,Kar:2020bws,Bernreuther:2020vhm,Cohen:2020afv,Lim:2020igi,Knapen:2021eip,Linthorne:2021oiz,Beauchesne:2019ato}.
In this paper we discuss another possibility for composite ADM model: mediator $\Phi$ couples to a dark quark and a lepton, instead of a dark quark and a SM quark. 
The rapid sphaleron process in the early universe can transfer the lepton asymmetry generated from $\Phi$ decay into baryon asymmetry, and thus BAU can also be explained in this model. 
However, different with the model proposed in~\cite{Bai:2013xga}, the dark sector in this model mainly talks to the lepton sector in the SM. 
Such a leptophilic composite ADM model can be detected by direct detection (DD) experiments through the scattering with electrons.
This model can also be detected at colliders by novel signatures like displaced lepton jets. 
We will choose LHC and projected CEPC~\cite{CEPCStudyGroup:2018ghi} as representative detectors for collider searches.  

In the next section, we will construct an ADM model which explain the BAU and DM relic density simultaneously. 
In section III we discuss the constrains on this model from DD experiments. 
Section IV is dedicated to the collider searches. 
In section V we give a brief discussion about recent muon g-2 results. 
In section VI we conclude this work. 

\section{The generation of baryon and dark baryon asymmetry \label{sec:Model} }

The model we present in this section is similar to the model proposed in~\cite{Bai:2013xga}. 
Firstly we need to generate the asymmetry of mediator $\Phi$. 
This goal can be achieved by the out-of-equilibrium \& CP violated decay of a heavy Majorana fermion, which is just like the leptogenesis process~\cite{Fukugita:1986hr,Buchmuller:2004nz,Davidson:2008bu}. 
Furthermore, we need to break the conservation of dark baryon number (labeled as $B'$) when the temperature of universe is high. 
In addition to the scalar mediator $\Phi$ and dark quark $q'$~\footnote{We can certainly consider more than one flavor of dark quark, but the number of dark quark flavor is not relevant in this section.}, we extend the SM by two heavy Majorana fermions ($N_1$ and $N_2$) for out-of-equilibrium process, and a Dirac fermion $\chi$ charged under dark QCD (denoted as $SU(3)'$) to break the dark baryon number $B'$. 
In Tab.~\ref{particle} we present charge, spin, and particle number carried by each particle.

\begin{table}[htp]
\begin{center}\begin{tabular}{|c|c|c|c|c|c|c|c|}
\hline   & $ \ SU(3)'  \ $ & $ \ SU(3) \ $ & $ \ U_Y(1) \ $ & \  Spin  \ & $ \ L \ $ & $ \ B \ $ & $ \ B' \ $ \\
\hline $N_1/N_2$ & 1   & 1  & 0  & 1/2  & 0  & 0  & 0   \\
\hline $\Phi$    & 3  & 1  & 1  & 0   &  -1 & 0  & 1/3  \\
\hline $\chi$ & 3  & 1  & 1  & 1/2  &  -1 & 0  & 1/3  \\
\hline $q'$       & 3  & 1  & 0  & 1/2  & 0  & 0  & 1/3  \\
\hline $l_R$    & 1 & 1  & -1  &  1/2 & 1  & 0 & 0  \\
\hline $d_R$    & 1 & 3  & -1/3  &  1/2 & 0  & 1/3 & 0  \\
\hline $u_R$    & 1 & 3  & 2/3  &  1/2 & 0  & 1/3 & 0  \\
\hline \end{tabular} 
\caption{Particles contents and their property. Here we present charges, spins, and particle numbers carried by each particles. 
$N_1/N_2$ are heavy Majorana fermions.
$\Phi$ is the scalar mediator that carries the SM lepton number $L$ and dark baryon number $B'$.  
$\chi$ is a Dirac fermion that carries the same particle number and charge as $\Phi$ carries. 
$q'$ is the dark quark. 
$l$ denote the SM leptons $e$, $\mu$, and $\tau$.
$d_R$ and $u_R$ are the SM right-handed down quark and up quark.}
\label{particle}
\end{center}
\end{table}

We extend the Lagrangian of the SM with the particles present in Tab.~\ref{particle}:
\begin{eqnarray}
\label{lagrangian}
\mathcal{L} = & & \mathcal{L}_{\text{SM}} -\frac{1}{2} \sum_{i=1,2} M_{N_i} \overline{N}_i N^C_i  -m^2_{\Phi} \Phi^{\dagger} \Phi - m_{\chi} \bar{\chi}\chi - m_{q'} \bar{q'}q'  + \mathcal{L}_{\text{kinetic}}        \\ \notag
   & & - \sum_{i=1,2} \lambda_i \overline{N}_i \chi \Phi^{\dagger} - \kappa \Phi \bar{q'}_L  {l}_R - \frac{1}{\Lambda_1^2} \left( \bar{q'}^C \chi  \right) \left( \bar{q'}^C_L l_R  \right) - \frac{1}{\Lambda_2^2} \left(\bar{\chi} \gamma^{\mu} q'  \right) \left( \bar{d}_R \gamma_{\mu} u_R  \right) + h.c. 
\end{eqnarray}
In the above Lagrangian, we dismiss the traditional leptogenesis operator $N_i L H$ by assuming this term to be negligible, because we don't want to mix the standard leptogenesis with our scenario. 
Furthermore, we introduce two dimension-6 four fermions interactions to make $\chi$ decay, and at the same time make the decay process  
break dark baryon number $B'$ or lepton number $L$~\footnote{A dimension-5 operator $(\chi \Phi^{\dagger})(LH)$ , which is obtained by integrating out heavy $N_i$, can also make $\chi$ decay. But due to the huge hierarchy between $M_{N_i}$ and $m_{\chi}$, we assume this operator to be negligible.}. 
Actually, due to angular momentum conservation and gauge invariance, 
we can not write down dimension-4 operator to make $\chi$ decay~\footnote{By introducing additional new particles and new interactions, we can design a cascade decay chain to make $\chi$ decay without high dimension operator. However, in this work we focus on the study of $\Phi$, and thus we simply use dimension-6 operators to make $\chi$ unstable.}. 
We choose $N_1$ to be lighter than $N_2$. 
One of CP phases of parameter $\lambda_i$ can be rotated away by field redefinition, and the rest one is the source of CP violation. 
Coupling parameter $\kappa$ and scale $\Lambda_{1,2}$ can be chosen to be different for different lepton flavor, but here we assume them to be lepton flavor universal for simplicity. 
For conciseness we do not write out the specific form of kinetic terms $\mathcal{L}_{\text{kinetic}}$.

In the following, we explain how to generate the baryon and dark baryon asymmetries in this model. 
For convenience, we use ``$Y$'' to denote the net number density of particle ``$i$'' relative to entropy density $s$. 
For self-conjugate particle and non-self-conjugate particle, expressions of $Y$ are slightly different:
\begin{eqnarray}
Y_{i} &\equiv& \frac{n_i}{s} \ \ \ \ \ \ \ \ \ \ (\text{if ``$i$'' is self-conjugate})  \\\nonumber
Y_{\Delta i} &\equiv& \frac{n_i - n_{\bar{i}}}{s}  \  \ \ \  (\text{if ``$i$'' is not self-conjugate})
\label{Y_def}
\end{eqnarray}

\subsection{Stage I: out-of-equilibrium \& CP violated decay of $N_1$}

In this work we consider following hierarchy of particle spectrum:
\begin{eqnarray}
M_{N_1},\ M_{N_2} \gg m_{\Phi}, \ m_{\chi} \gg m_{q'}, \ m_{l}
\end{eqnarray}
For simplicity we assume $N_1$ is in the thermal equilibrium when universe temperature $T \gg M_{N_1}$. 
At that temperature, the number density of $N_1$ is quantified as:
\begin{eqnarray}
Y_{N_1} = \frac{n_{N_1}}{s} = \frac{135 \zeta(3)}{4\pi^4 g_{\ast}}
\end{eqnarray}
Here $\zeta(3)\simeq 1.2$, and $g_{\ast}$ is the number of relativistic degree of freedom, which can be approximated to 100. 
As the universe temperature decreases, $N_1$ starts to decay. 
If this decay process is out-of-equilibrium and $\lambda_i$ have irreducible CP phases, asymmetry between the number density of $\Phi$/$\chi$ and their anti-particles can be generated~\cite{Buchmuller:2004nz,Davidson:2008bu}:
\begin{eqnarray}
Y_{\Delta \Phi} = - Y_{\Delta \chi} = Y_{N_1}\times \epsilon \times \eta
\label{YN1}
\end{eqnarray}

$\epsilon$ in Eq.~\ref{YN1} is the parameter used to describe the CP asymmetry in $N_1$ decay:
\begin{eqnarray}
\epsilon \equiv \frac{\Gamma(N_1 \to \chi \Phi^{\dagger}) - \Gamma(N_1 \to \bar{\chi} \Phi)}{\Gamma(N_1 \to \chi \Phi^{\dagger}) + \Gamma(N_1 \to \bar{\chi} \Phi)}
\end{eqnarray}
To obtain a non-zero value of $\epsilon$, the interference between tree level and one-loop level decay amplitudes needs to be calculated. 
In the case where $M_{N_2}$ is much larger than $M_{N_1}$, $\epsilon$ can be simply expressed as~\cite{Covi:1996wh}:
\begin{eqnarray}
\epsilon \simeq -\frac{3}{16\pi} \frac{M_{N_1}}{M_{N_2}} \frac{ \text{Im} \left[ (\lambda^{\ast}_2 \lambda_1 )^2 \right] }{|\lambda_1|^2}
\label{asy}
\end{eqnarray}
Depending on the sign of $\text{Im} \left[ (\lambda^{\ast}_2 \lambda_1 )^2 \right]$, $\epsilon$ can be positive or negative. 

$\eta$ in Eq.~\ref{YN1} is the efficiency factor that describe the ``washout'' effects. 
Asymmetries generated in Eq.~\ref{asy} can be erased by processes like inverse decay ($\chi+\Phi^{\dagger}\to N_1$) or 2-to-2 scattering ($ \bar{q'}+l_R \to N_1+ \bar{\chi}$).
If $N_1$ is always in the thermal equilibrium, then $\eta$ will be zero. 
Here we can choose the parameter $\lambda_1$ to be small enough, to make the decay width of $N_1$ be much smaller than the Hubble expansion rate when temperature is around $M_{N_1}$:
\begin{eqnarray}
\Gamma_{N_1} \ll H(T \approx M_{N_1})
\label{washout}
\end{eqnarray}
We can always choose suitable values for $\lambda_1$ and $M_{N_1}$ that satisfy Eq.~(\ref{washout}) and make $N_1$ in the thermal equilibrium when $T \gg M_{N_1}$.
In this case, the value of $\eta$ can be close to 1.

Before moving to the next stage, here we estimate the scale of $|Y_{\Delta \Phi}|$ and $|Y_{\Delta \chi}|$. 
$Y_{N_1}$ is about $4\times 10^{-3}$. 
If we choose $\lambda_1$ to be real, $\text{Im} (\lambda^{\ast}_2)^2 $ to be $\mathcal{O}(1)$, and $\frac{M_{N_1}}{M_{N_2}}$ to be $\mathcal{O}(0.01)$,
then $|Y_{\Delta \Phi}|$ and $|Y_{\Delta \chi}|$ can be as large as $2\times 10^{-6}$. 
Smaller $|Y_{\Delta \Phi}|$ and $|Y_{\Delta \chi}|$ can be easily obtained by decreasing the value of $\text{Im} (\lambda^{\ast}_2)^2 $ or $\frac{M_{N_1}}{M_{N_2}}$.

\subsection{Stage II: $\Phi$ and $\chi$ decay to leptons and dark quarks}

In the Lagrangian we introduce two operators of $\chi$ to break $B'$ and $L$ conservation. 
$\left( \bar{q'}^C \chi  \right) \left( \bar{q'}^C_L l_R  \right)$ breaks $B'$ number by 1, and $\left(\bar{\chi} \gamma^{\mu} q'  \right) \left( \bar{d}_R \gamma_{\mu} u_R  \right)$ breaks $L$ number by 1. 
Both operators are necessary for the generation of baryon and dark baryon asymmetries. 
This is because $\Phi$ and $\chi$ carry the same particle number, but after $N_1$ decay we obtain $Y_{\Delta \Phi} = - Y_{\Delta \chi}$. 
So the net particle number density of $q'$ and $l$ can not be generated without $B'$ and $L$ violated processes. 
We denote the branching ratios of decay channels $\chi\to\bar{q'}{q'}^C\bar{l}_R$ and $\chi\to{q'}\bar{d}_R u_R$, by $Br_{\chi}(B' \!\!\!\!\!/ \ )$ and $Br_{\chi}(L \!\!\!/)$ respectively. 
Because $\chi$ only has two decay channels, so we should have $Br_{\chi}(B' \!\!\!\!\!/ \ )+Br_{\chi}(L \!\!\!/) = 1$.

We need the Sphaleron process in the SM to transfer lepton number to baryon number. 
Lattice simulation shows that the SM Sphaleron transition rate will be smaller than the Hubble expansion rate when the universe temperature is $T_{\ast} =  (131.7\pm 2.3) \text{ GeV}$~\cite{DOnofrio:2014rug}. 
So $\Phi$ and $\chi$ need to decay before $T_{\ast}$.
This requirement can be satisfied by choosing $m_{\Phi}$ and $m_{\chi}$ to be higher than $T_{\ast}$, and make their lifetime shorter than $1/H(T_{\ast})$. 
If we assume that the net number densities of $\Phi$ and $\chi$ have been almost entirely transferred to the number densities of $l_R$ and $q'$ before $T_{\ast}$. 
Then there will be\footnote{Here we also require decay processes $\chi\to\bar{q'}{q'}^C\bar{l}_R$ and $\chi\to{q'}\bar{d}_R u_R$ to be out-of-equilibrium. Otherwise the decay products of these two processes will transfer to each other, through $ \bar{q'}{q'}^C\bar{l}_R \leftrightarrow \chi \leftrightarrow {q'}\bar{d}_R u_R$. And thus the estimation of particle number density will be more complicated.}: 
\begin{eqnarray}
Y_{\Delta l_R} &\simeq& - Y_{\Delta \Phi} -  Br_{\chi}(B' \!\!\!\!\!/ \ ) \times Y_{\Delta \chi} = - Br_{\chi}(L \!\!\!/) \times Y_{\Delta \Phi}    \\
Y_{\Delta q'} &\simeq& Y_{\Delta \Phi} + ( Br_{\chi}(L \!\!\!/) - 2 Br_{\chi}(B' \!\!\!\!\!/ \ ) ) \times Y_{\Delta \chi} = 3 Br_{\chi}(B' \!\!\!\!\!/ \ ) \times Y_{\Delta \Phi}
\end{eqnarray}

\subsection{Stage III: generate baryon and dark baryon asymmetries}

\begin{figure}[ht]
\centering
\includegraphics[width=6.7in]{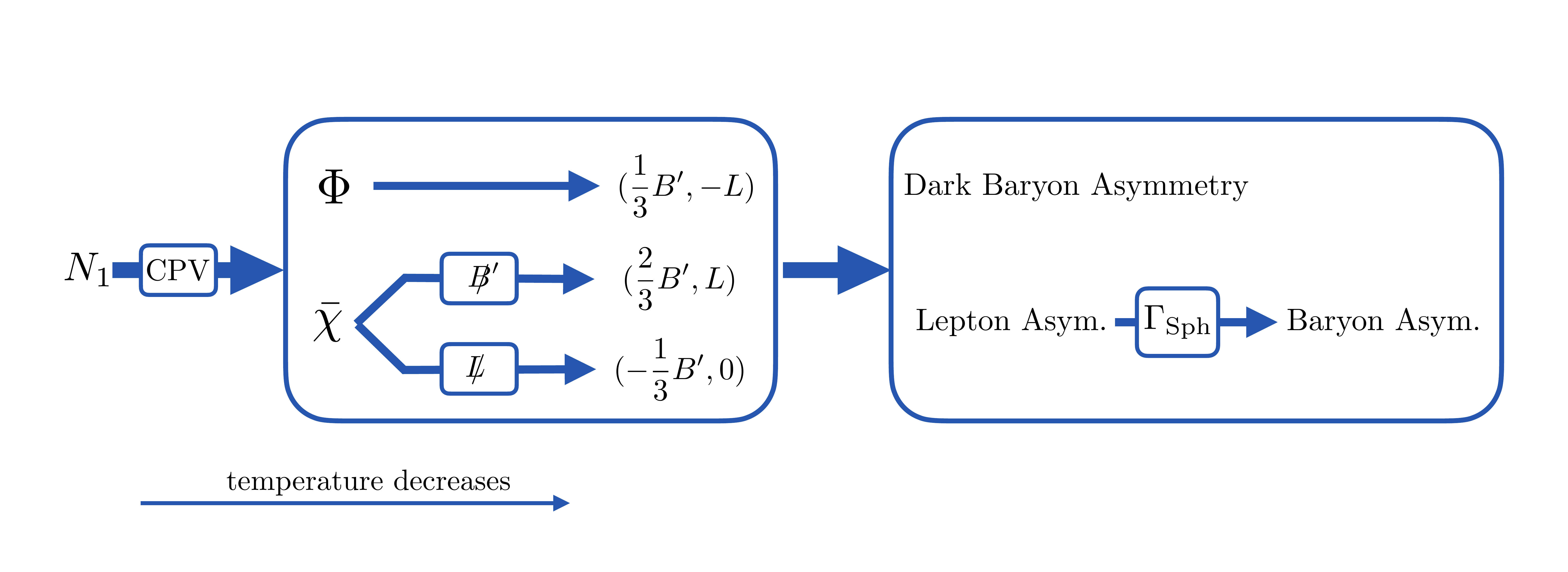}
\caption{A schematic diagram for the DM asymmetry and baryon asymmetry generation process. 
Firstly, the out-of-equilibrium \& CP violated decay of $N_1$ generates the asymmetries of $\Phi$ and $\chi$. 
Then the decay of $\Phi$ and $\chi$ generate the asymmetries of dark baryon and lepton. 
Finally, the lepton asymmetry is transferred to baryon asymmetry via sphaleron process. 
}
\label{history}
\end{figure}

Each dark quark $q'$ carries $1/3$ dark baryon number, thus after $\Phi$ and $\chi$ decay we obtain:
\begin{eqnarray}
Y_{\Delta B'} = \frac{1}{3} Y_{\Delta q'} \simeq Br_{\chi}(B' \!\!\!\!\!/ \ ) \times Y_{\Delta \Phi} 
\end{eqnarray}
On the other hand, gauge interaction, Yukawa interaction, and Sphaleron transition in the SM are rapid when the universe temperature is higher than $T_{\ast}$. 
A combination of all these processes finally transfer the lepton number, which are carried by $l_R$, to baryon number~\cite{Harvey:1990qw}: 
\begin{eqnarray}
Y_{\Delta B} = \frac{28}{79} Y_{\Delta B-L} = - \frac{28}{79} Y_{\Delta l_R} \simeq  \frac{28}{79} Br_{\chi}(L \!\!\!/) \times Y_{\Delta \Phi}
\end{eqnarray}
As explained in stage I,  $|Y_{\Delta \Phi}|$ can be as large as $2\times 10^{-6}$, 
and thus it is easy to explain the observed BAU $Y_{\Delta B}  \approx 8\times 10^{-11}$, provided $Br_{\chi}(L \!\!\!/)$ is not too small.

The mass of dark baryon can be fixed by the ratio between $Y_{\Delta B'}$ and $Y_{\Delta B}$:
\begin{eqnarray}
\frac{m_{B'} Y_{\Delta B'} }{m_{B} Y_{\Delta B}} = \frac{\Omega_\text{DM}}{\Omega_B} \ \Rightarrow \ m_{B'} \simeq \frac{Br_{\chi}(L \!\!\!/)}{Br_{\chi}(B' \!\!\!\!\!/ \ )} \times 1.67 \text{ GeV}
\end{eqnarray}
So in this model, dark baryon mass relies on the branching ratios of $\chi$, and can vary in a wide range. 
In Fig.~\ref{history} we present a schematic diagram to summarizes above asymmetry generation process. 

\subsection{Stability of dark baryon $B'$ and dark pion $\pi'$}

\begin{figure}[ht]
\centering
\subfigure{
\includegraphics[width=3.0in]{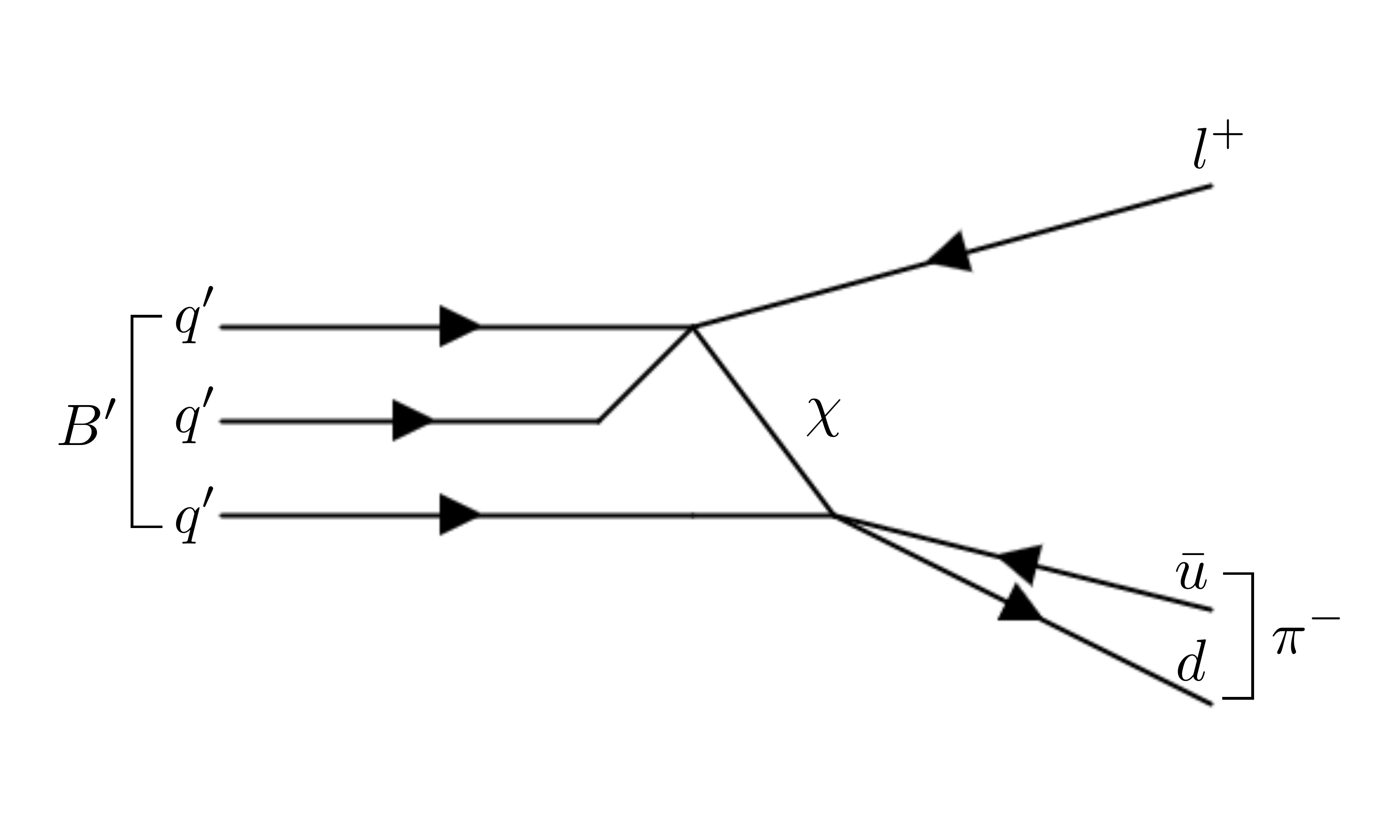}
}
\quad
\subfigure{
\includegraphics[width=3.0in]{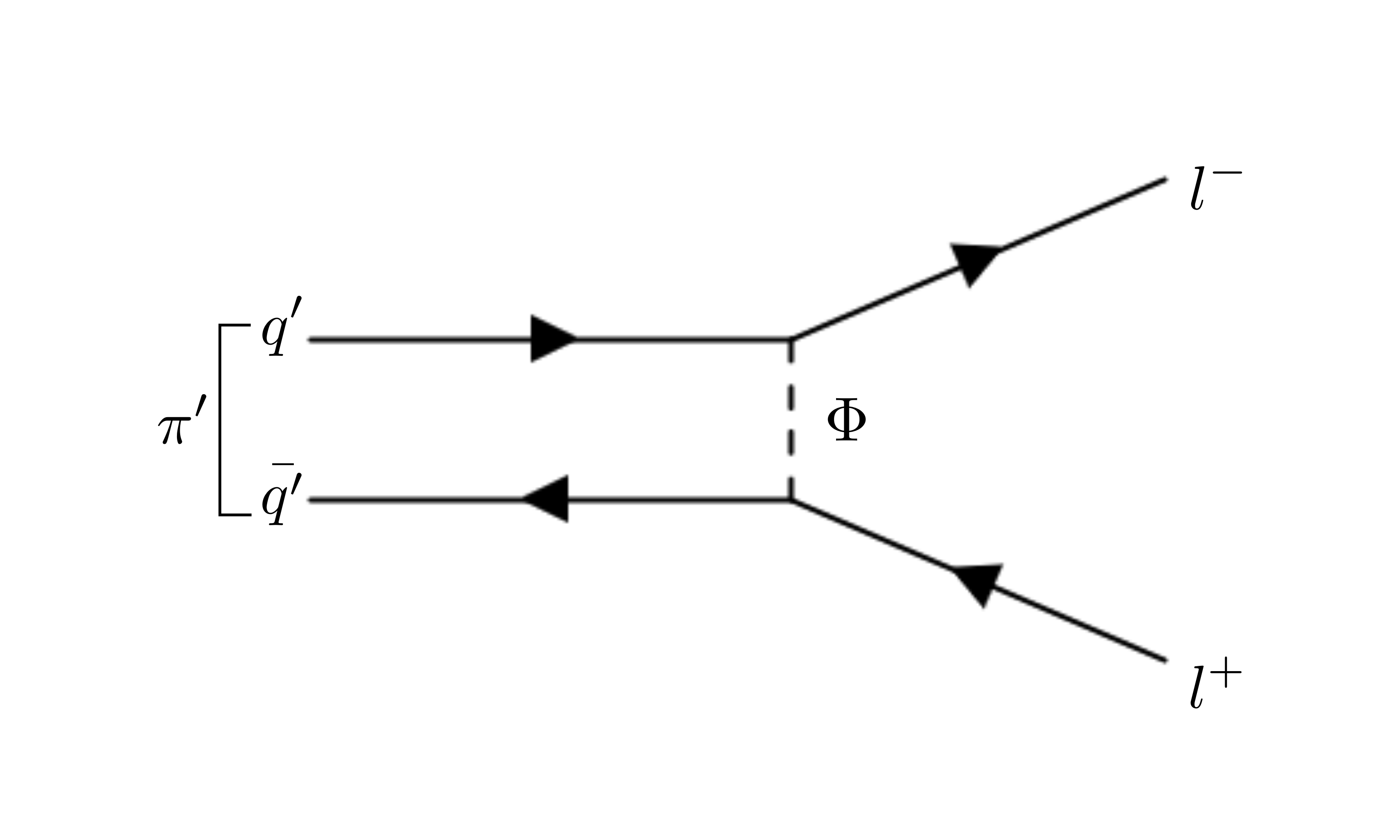}
}
\caption{Left: illustration for dark baryon $B'$ decaying to $l^+  \pi^-$ via $\chi$. Right: illustration for dark pion $\pi'$ decaying to $l^+  l^-$ via $\Phi$.}
\label{decay}
\end{figure}

$\chi$ in Lagrangian~(\ref{lagrangian}) couples to a $B'$ violated operator and a $L$ violated operator, and thus a dark baryon (composed by $q'q'q'$) can decay to $l^+  \pi^-$ via $\chi.$ 
One the other hand, dark pion (composed by $\bar{q'}q'$) can decay to $l^+  l^-$ via a t-channel $\Phi$.
In Fig.~\ref{decay} we present these two decay processes for illustration. 

By integrating out $\chi$ and $\Phi$ in the Lagrangian, we obtain a dimension-9 operator and a dimension-6 operator that induce $B'$ and $\pi'$ decay respectively:
\begin{eqnarray}
\mathcal{L} \supset \frac{\kappa^2}{m^2_{\Phi}} \left( \bar{q'}_L  {l}_R  \right) \left( \bar{l}_R  {q'}_L  \right) + \left[ \frac{1}{m_{\chi} \Lambda_1^2 \Lambda_2^2} \left( \bar{q'}^C_L l_R  \right) \left(\bar{q'}^C \gamma^{\mu} q'  \right) \left( \bar{d}_R \gamma_{\mu} u_R  \right) + h.c. \right]
\label{eft}
\end{eqnarray}
Thus the decay width of $B'$ is suppressed by the factor $m^2_{\chi} \Lambda_1^4 \Lambda_2^4$. 
Current bound on dark matter lifetime from weak lensing and cluster counts is $\tau_{B'} \gtrsim 175 $ Gyr~\cite{Enqvist:2019tsa}. 
This bound can be rewritten as $\Gamma(B') \lesssim 1.2 \times 10^{-43}$ GeV. 
To satisfy this lifetime bound, we can simply put $m_{\chi}$ and $\Lambda_{1,2}$ to a high scale which is unreachable for current particle physics experiments.  
So in the following sections we will not discuss the detection of $\chi$.


In our ADM scenario, anti-DM $\bar{B'}$ in the early universe will be almost completely wiped out by the annihilation with DM $B'$. 
And thus the annihilation process $B' \bar{B'}\to \pi' \pi'$ produce plenty of dark meson $\pi'$ in the early universe. 
If those $\pi'$ are long-lived like $B'$, they will overclose the Universe (if $\pi'$ is heavy) or affect the effective number of neutrino degrees of freedom $N_\text{eff}$ (if $\pi'$ is light)~\cite{Blennow:2012de}. 
So $\pi'$ should decay before BBN.
$\pi'$ decays to lepton pair through the dimension-6 operator in Lagrangian~\ref{eft}.
Its decay width to lepton pair $\bar{l}l$ is~\cite{Schwaller:2015gea}:
\begin{eqnarray}
\Gamma(\pi' \to \bar{l}l) &=& \frac{\kappa^4}{32\pi m^4_{\Phi}} f^2_{\pi'} m^2_{l} m_{\pi'} \ , \ \text{with } l = e, \mu, \tau
\end{eqnarray}
Here $f_{\pi'}$ is the decay constant of dark pion $\pi'$. 
Due to the hierarchy between $m_e$, $m_\mu$, and $m_\tau$, the lifetime of $\pi'$ is determined by the decay width to the heaviest lepton pair allowed by kinematics. 
The constraint from BBN give us a bound on $\pi'$ lifetime: 
\begin{eqnarray}
\frac{1}{\Gamma(\pi' \to \bar{l}l)} \lesssim 1.52\times 10^{24} \text{ GeV}^{-1} \ \Rightarrow m_{\Phi} \lesssim 0.35\times 10^6 \times \kappa\sqrt{ f_{\pi'} m_l  } \times \left( \frac{m_{\pi'}}{1\text{ GeV}} \right)^{\frac{1}{4}}
\end{eqnarray}
If we choose a benchmark setting like: $\kappa = 1, \ f_{\pi'} = m_{\pi'} = 0.3\text{ GeV}.$ Then $\pi'$ can only decay to $\bar{\mu}\mu$ and the up-limit of $m_{\Phi}$ from BBN  is $2.6\times 10^5$ GeV.
If the dark confinement scale $\Lambda'$ become higher and $m_{\pi'}$ can decay to $\bar{\tau}\tau$, then the upper limit of $m_{\Phi}$ can be increased by more than an order of magnitude.

One the other hand, $\Phi$ can not be lighter than the Sphaleron freeze-out temperature $T_{\ast} \approx 131.7 \text{ GeV}$, because we need to generate enough anti-leptons via $\Phi$ decay, before Sphaleron process stop. 
We assume the coupling $\kappa$ for interaction $ \Phi \bar{q'}_L  {l}_R$ is not too small, and thus $\Phi$ is always in the thermal bath during its decay. 
Then the proportion of $\Phi$ particles that have decayed before $T_{\ast}$ can be estimated by $1-e^{-m_{\Phi}/T_{\ast}}$. 
So even $\Phi$ is as light as 300 GeV, 90\% of $\Phi$ have decayed before Sphaleron process stop. 
This is enough to explain BAU in a large parameter space of this model. 

Such a small mass lower-limit makes scalar mediator $\Phi$ a promising target for current and coming particle physics experiments. 
In the following sections, we will discuss direct detection and collider detection of this model, via the portal provided by scalar mediator $\Phi$.
To simplify our analysis, in this work we will assume that $Br_{\chi}(L \!\!\!/)$ and $Br_{\chi}(B' \!\!\!\!\!/ \ )$ are equal to each other, and fix $m_{B'}$ to 2 GeV. 
This choice also fix the value of dark confinement scale: $\Lambda'_{\text{QCD}} \simeq \frac{m_{B'}}{m_{B}} \Lambda_{\text{QCD}} \simeq 0.64 \text{ GeV}$.

\section{direct detection}\label{sec:dd}

In this section we discuss the direct detection of this model. 
Dark baryon $B'$, which is composed by 3 dark quarks, can scatter with electrons in atoms through the dimension-6 operator:
\begin{eqnarray}
\frac{\kappa^2}{m^2_{\Phi}} \left( \bar{q'}_L  {e}_R  \right) \left( \bar{e}_R  {q'}_L  \right) = \frac{\kappa^2}{2 m^2_{\Phi}} \left( \bar{q'}_L \gamma^{\mu} {q'}_L   \right) \left( \bar{e}_R \gamma_{\mu}  {e}_R  \right)
\end{eqnarray}
Here we use Fierz identity to rewrite the operator.
So it is possible to detect $B'$ by direct detection experiments via electron ionization.

The matrix element for the dominant spin-independent scattering is~\cite{Lin:2019uvt}:
\begin{eqnarray}
\mathcal{M} = \frac{\kappa^2}{8 m^2_{\Phi}} g_{\mu\nu} J_{B'}^{\mu} J_e^{\nu}
\end{eqnarray}
where $J_e^{\nu} = \bar{u}(p') \gamma^{\nu} u(p)$, and $J_{B'}^{\mu} = \left\langle B'(k')\left|\bar{q'} \gamma^{\mu} q'\right| B'(k)\right\rangle \approx 3 \bar{u}(k') \gamma^{\mu} u(k)$, in the non-relativistic limit. 
Here $k$ ($p$) and $k'$ ($p'$) are initial and final state momentum of $B'$ (electron) respectively. 
So the spin-independent cross section of $B'$-electron scattering is:
\begin{eqnarray}
\bar{\sigma}_{eB'} \approx \frac{9 \kappa^4 \mu^2_{eB'}}{64 \pi m^4_{\Phi}}
\end{eqnarray}
Here $\mu_{eB'} = (m_e m_{B'} )/(m_e + m_{B'})$ is the reduced mass. 
If we choose $\kappa = 1$ and $ m_{\Phi} = 300$ GeV, then $\bar{\sigma}_{eB'} \approx 5.6\times 10^{-46}$ cm$^{2}$.
This is much smaller than current direct detection limits from XENON100~\cite{Essig:2017kqs,Aprile:2016wwo} or DarkSide-50~\cite{Agnes:2018oej}.
On the other hand, dark baryon $B'$ can also scatter with nucleons via loop induced process. 
But the corresponding DM-nucleon scattering cross section will be suppressed by loop factor. 
Furthermore, direct detection bounds on DM-nucleon scattering sharply become loose when DM mass is lighter than 10 GeV. 
So here we conclude that current direct detection experiments could hardly constrain our model, provided $m_{B'}$ is around GeV scale.

\section{Collider Search}\label{sec:collider}

In this section we study the detection of this model by collider experiments, including both hadron collider (e.g. LHC) and lepton collider (e.g. CEPC). 
The Lagrangian which is related to collider search can be expressed as~\cite{Schwaller:2015gea}:
\begin{eqnarray}
\mathcal{L} \supset \bar{q'} ( D \!\!\!\!/  - m_{q'} ) q'  + (D_{\mu} \Phi )^{\dagger}(D^{\mu} \Phi ) - m^2_{\Phi} \Phi^{\dagger} \Phi -\frac{1}{4} {G'}^{\mu\nu} {G'}_{\mu\nu} - ( \kappa \Phi \bar{q'}_L  {l}_R  +h.c.)
\label{Lag_collider}
\end{eqnarray}
Here ${G'}^{\mu\nu}$ is the field strength of dark gluon. 
If dark quark $q'$ can be produced at colliders via dimension-6 operator or $\Phi$ decay, 
then generally it will carry an energy which is much larger than the dark confinement scale $\Lambda'_{\text{QCD}}$ (in Sec. II D we have fixed $\Lambda'_{\text{QCD}}$ to 0.64 GeV). 
Thus the energetic $q'$ will shower then evolve to a bunch of collinear dark mesons after hadronization. 
Such an object is often call ``dark jet''. 
Dark jet is not totally invisible at collider since dark meson can decay back to SM particles through some portal. 
As we have explained in Sec. II D, dark meson $\pi'$ can decay to lepton pair via the dimension-6 operator provided by $\Phi$. 
Here we estimate how long it can propagate before it decay.
Proper lifetime of $\pi'$ is~\cite{Schwaller:2015gea}:
\begin{eqnarray}
c\tau_0 = \frac{c \hbar }{\Gamma_{\pi'}} \approx 120 \text{ mm} \times \frac{1}{\kappa^4} \left( \frac{ 1 \text{ GeV} }{f_{\pi'}} \right)^2
\left( \frac{ 0.1\text{ GeV} }{m_l} \right)^2  \left( \frac{ 1\text{ GeV} }{m_{\pi'}} \right)   \left( \frac{ m_{\Phi} }{ 500 \text{ GeV} } \right)^4
\end{eqnarray}
So we can expect that, in a large parameter space of this model, dark mesons inside dark jet can propagate for a moment, and then decay to lepton pair at a place away from the primary vertex. 
Such a novel ``displaced lepton jet'' is the main signal for the collider searches of this model. 

To simplify our collider analysis, we fix the mass of $\pi'$ to 0.3 GeV, and thus the dominant decay channel of $\pi'$ is $\pi'\to \bar{\mu}\mu$.
Furthermore, we choose the proper lifetime of $\pi'$, $c\tau_0$, to be input parameter instead of $f_{\pi'}$.
Because $c\tau_0$ is directly related to collider phenomenology.  
$\kappa$ and $m_{\Phi}$ are other two input parameters for collider search.

For Monte Carlo simulation, we use {\sc FeynRules}~\cite{Alloul:2013bka} to write  Lagrangian~\ref{Lag_collider} into an UFO model file~\cite{Degrande:2011ua}. 
Parton level events are generated by  {\sc MadGraph\,5}~\cite{Alwall:2011uj}, then showered and hadronized by {\sc Pythia\,8}~\cite{Sjostrand:2014zea}. 
The {\sc HiddenValley}~\cite{Strassler:2006im} module implemented in {\sc Pythia\,8} can be used to simulate dark shower and dark hadronization process. 
Detector simulation is performed by {\sc DELPHES\,3}~\cite{deFavereau:2013fsa}. 
Finally, we use the anti-kt algorithm~\cite{Cacciari:2008gp} implemented in {\sc Fastjet}~\cite{Cacciari:2011ma} to do jet clustering if needed. 

\subsection{Detection at LHC}

ATLAS group already performed the displaced lepton jets signal search at 13 TeV LHC by using an event sample of integrated luminosity 3.4 fb$^{-1}$~\cite{ATLAS:2016jza}.
Deviations from the SM expectations are not observed. 
In this subsection we will use the results presented in~\cite{ATLAS:2016jza} to constrain our model.

\begin{figure}[ht]
\centering
\includegraphics[width=4.0in]{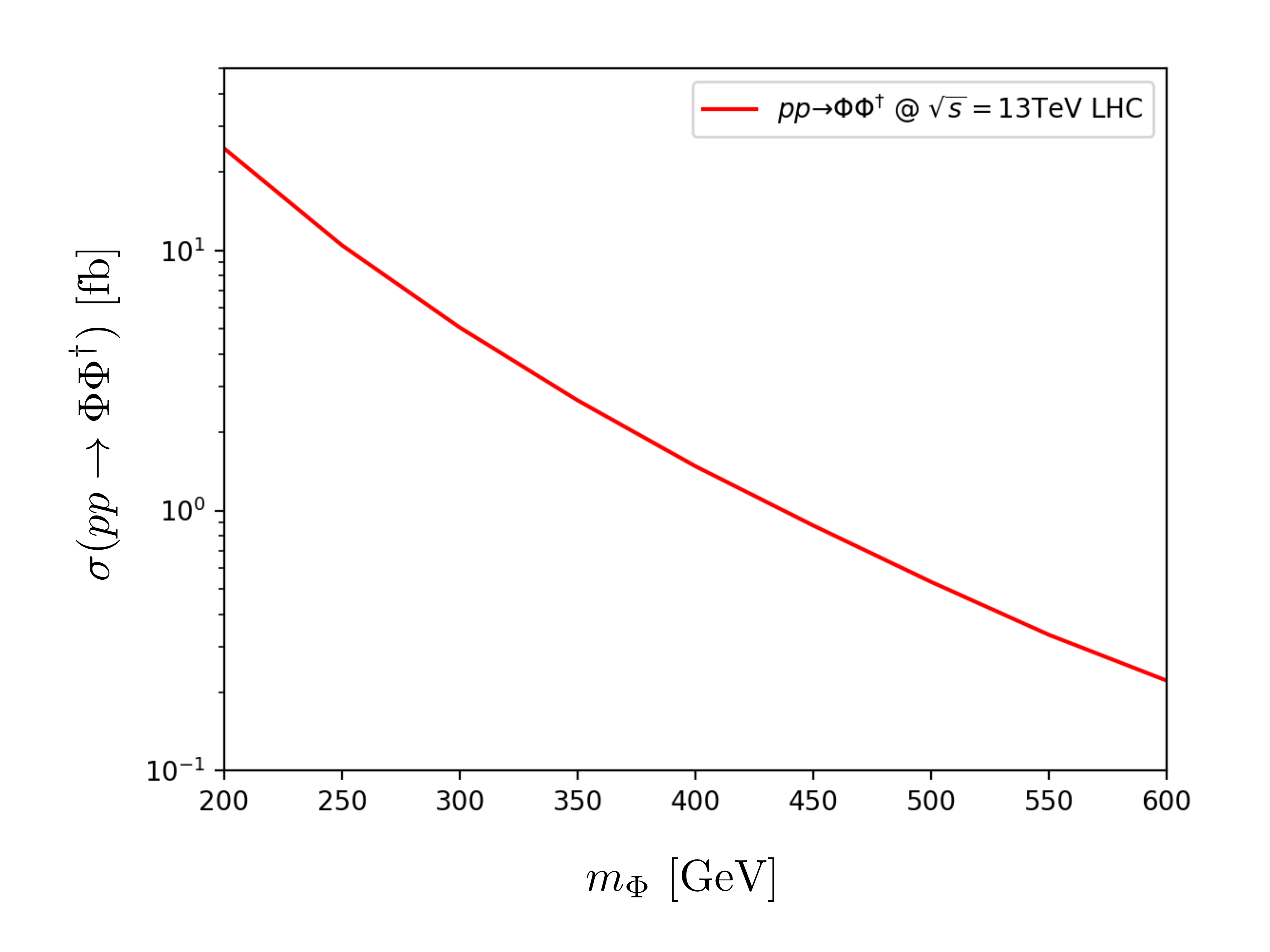}
\caption{Cross section of $\Phi$ pair production process at 13 TeV LHC as a function of $m_{\Phi}$. }
\label{LHC}
\end{figure}

Scalar mediator $\Phi$ carries Hyper charge $+1$ and dark $SU(3)'$ charge $3$.  
Thus the cross section of $\Phi$ pair production is just three times that of right-handed slepton pair production, when their masses are the same. 
NLO+NLL cross section of slepton pair production at 13 TeV LHC have been given in~\cite{Fiaschi:2018xdm}.
Based on their results, in Fig.~\ref{LHC} we present the cross section of $\Phi$ pair production at 13 TeV LHC. 
It can be seen that this cross section is quite small. 
For example, if $m_{\Phi}$ is 300 GeV, then integrated luminosity 3.4 fb$^{-1}$ can only generate 17.6 signal events. 
Event preselection in~\cite{ATLAS:2016jza} will eliminate more than half of the signal events. 
After that, depending on the position where long-lived particle decay, the tagging efficiency of displaced lepton jets varies from 50\% to 10\%.  
And the signal region require two displaced lepton jets.
Thus the original number of signal events will be suppressed by about an order, after performing the full cut flow in~\cite{ATLAS:2016jza}. 
On the other hand, the irreducible background for displaced muon jets, which mainly comes from cosmic-rays, is expected to be $31.8\pm 3.8$ (stat) $\pm 8.6$ (syst).
And experiment~\cite{ATLAS:2016jza} finally record 46 events.
So we can conclude that our model is very difficult to be detected or excluded by current LHC data.

\subsection{Detection at lepton collider}

\begin{figure}[ht]
\centering
\subfigure{
\includegraphics[width=3.0in]{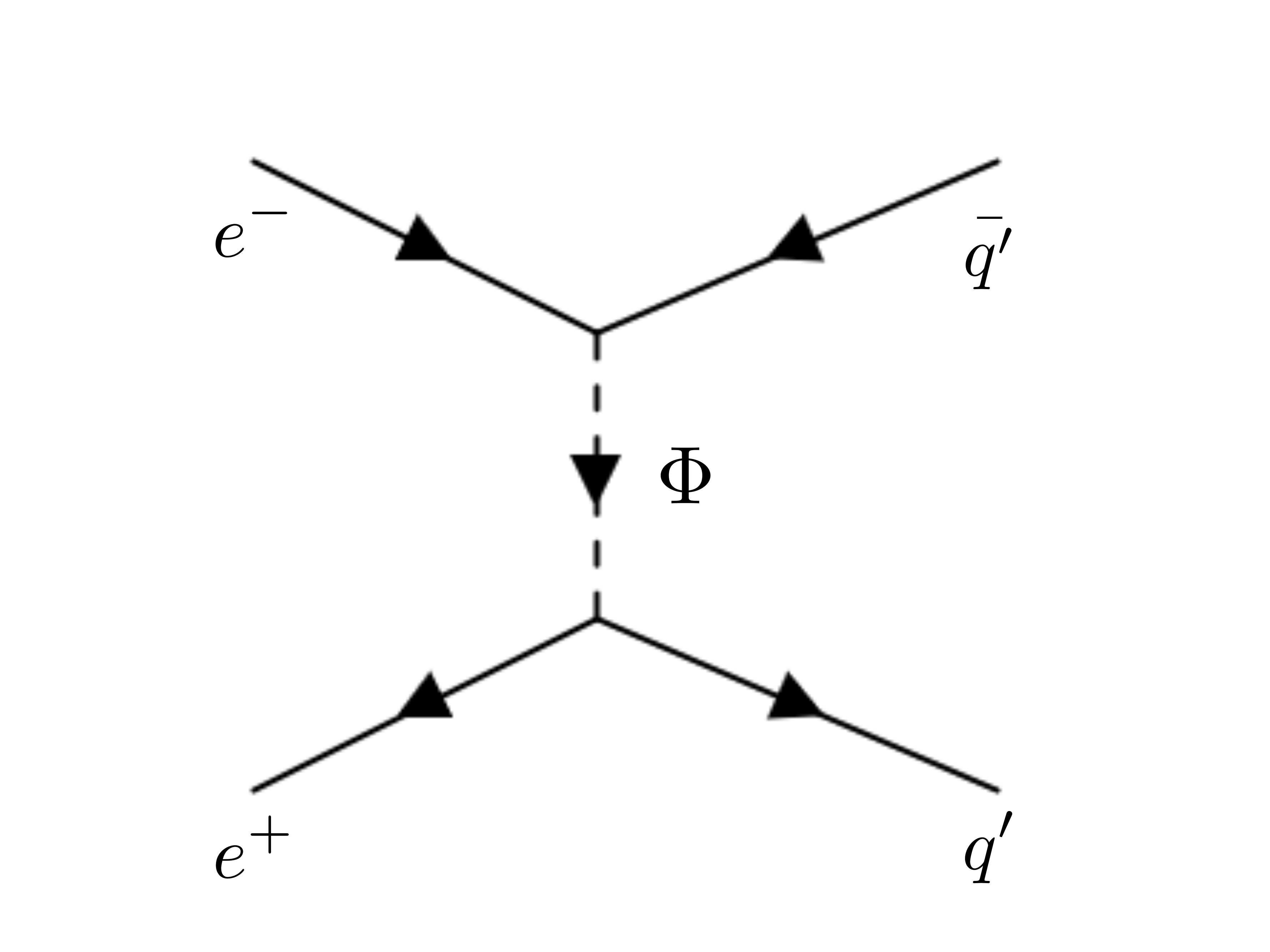}
}
\quad
\subfigure{
\includegraphics[width=3.0in]{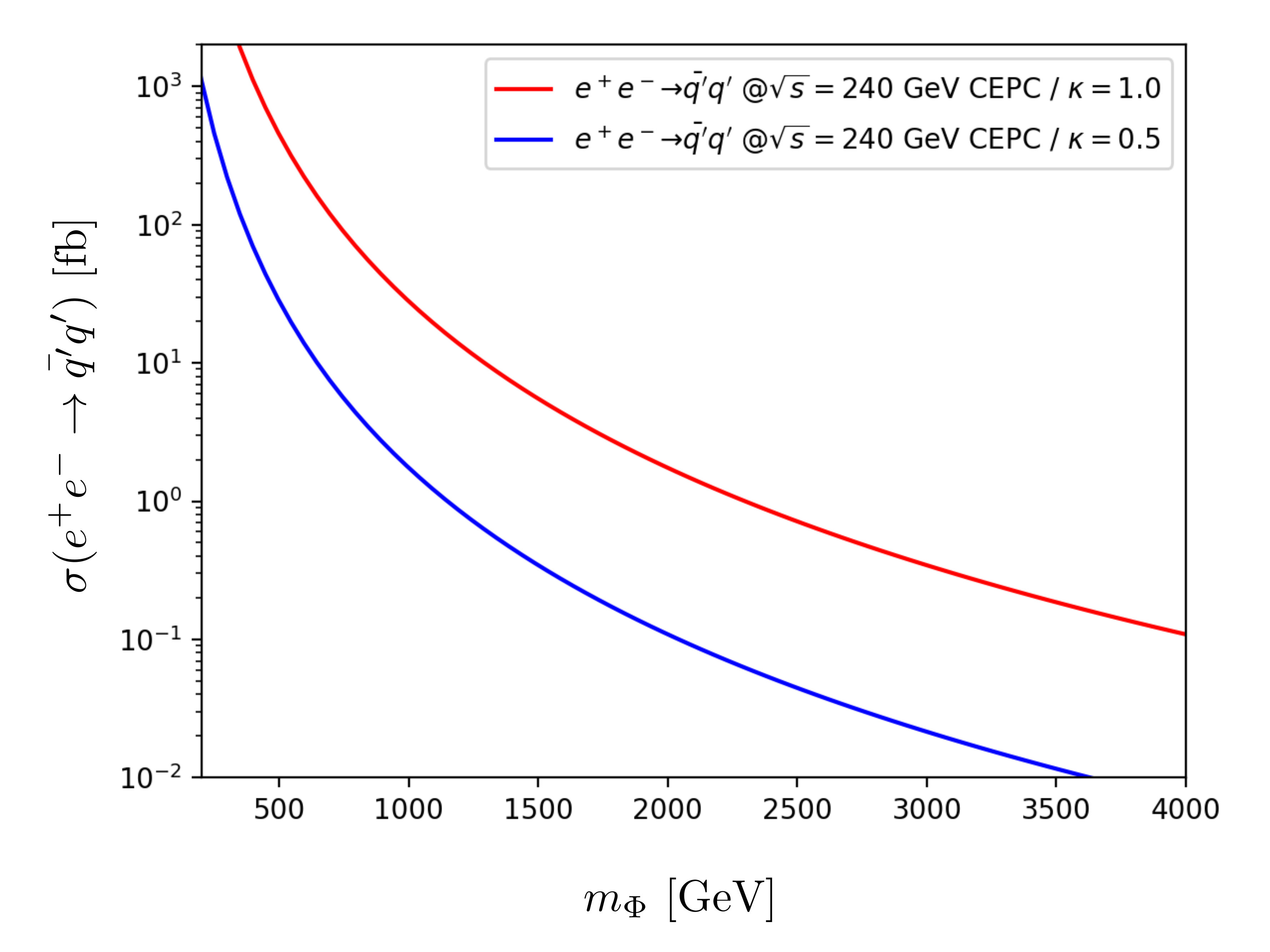}
}
\caption{Left: Feynman diagram of the signal process at CEPC. 
Right: Cross section of $q'$ pair production process at CEPC as functions of $m_{\Phi}$, with coupling $\kappa$ fixed to 0.5 and 1.0 respectively.}
\label{CEPC1}
\end{figure}

Different with hadron collider, projected lepton collides, including ILC~\cite{Baer:2013cma}, CLIC~\cite{Abramowicz:2013tzc}, FCC-ee~\cite{Gomez-Ceballos:2013zzn}, and CEPC~\cite{CEPCStudyGroup:2018ghi}, use electron and positron as injecting beam. 
Thus it is possible to directly generate dark quark pair through a t-channel process, see Fig.~\ref{CEPC1} (left) for illustration. 
When the mass of mediator $\Phi$ is large enough, we can integrate $\Phi$ and obtain an dimension-6 operator:
\begin{eqnarray}
\frac{\kappa^2}{m^2_{\Phi}} \left( \bar{q'}_L  {e}_R  \right) \left( \bar{e}_R  {q'}_L  \right) 
\end{eqnarray}
Thus the cross section of $e^{+}e^{-}\to \bar{q'}q'$ can be approximated as:
\begin{eqnarray}
\sigma(e^{+}e^{-}\to \bar{q'}q') \approx \frac{\kappa^4}{256\pi} \frac{s}{m^4_{\Phi}}
\end{eqnarray}
where electron and $q'$ are treated as massless. 
In the rest part of this subsection, we will choose CEPC, with central energy $\sqrt{s}=240$ GeV and integrated luminosity 5.6 ab$^{-1}$, as a benchmark setting for lepton collider detection. 

\begin{figure}[ht]
\centering
\subfigure{
\includegraphics[width=3.0in]{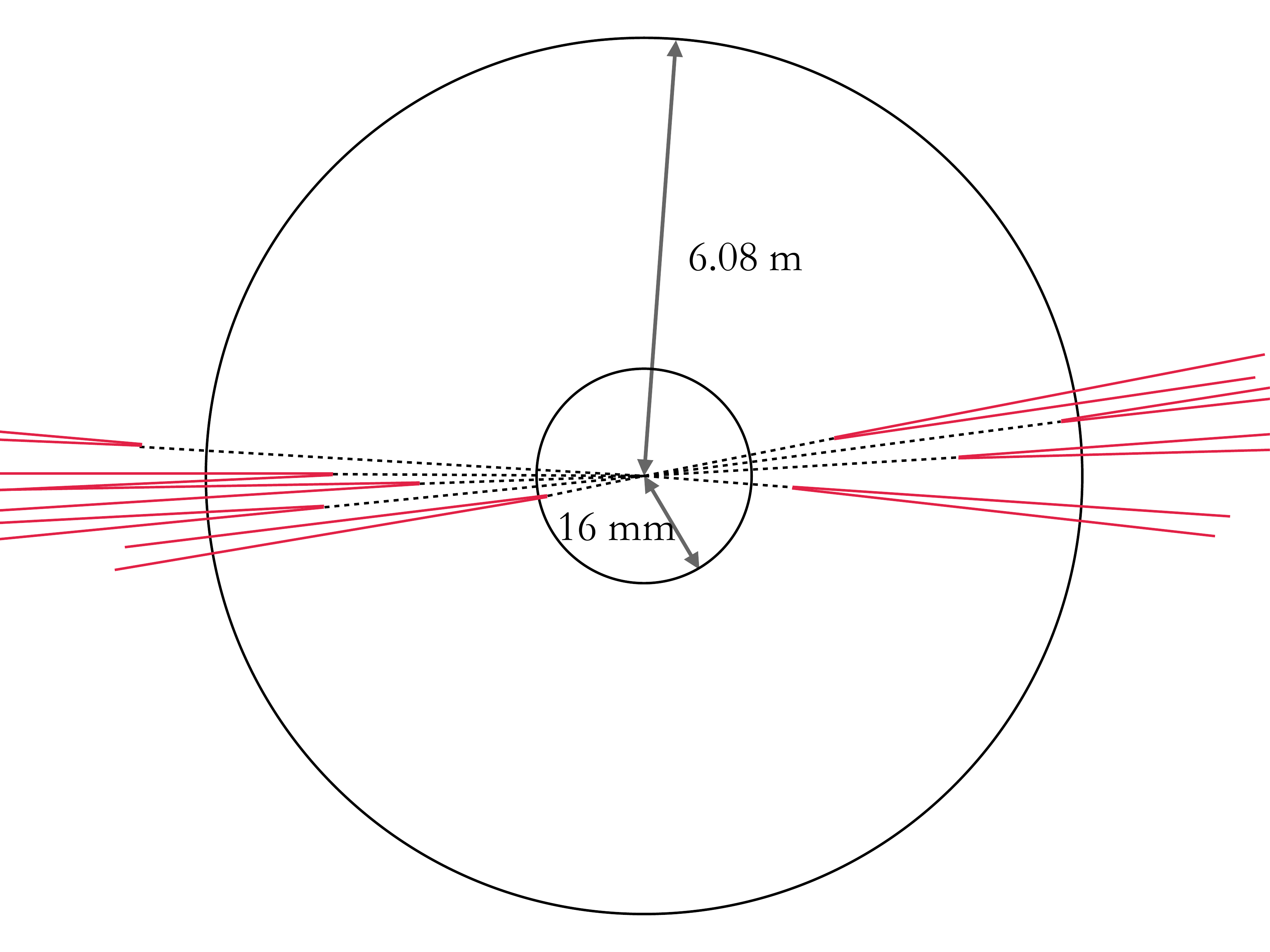}
}
\quad
\subfigure{
\includegraphics[width=3.0in]{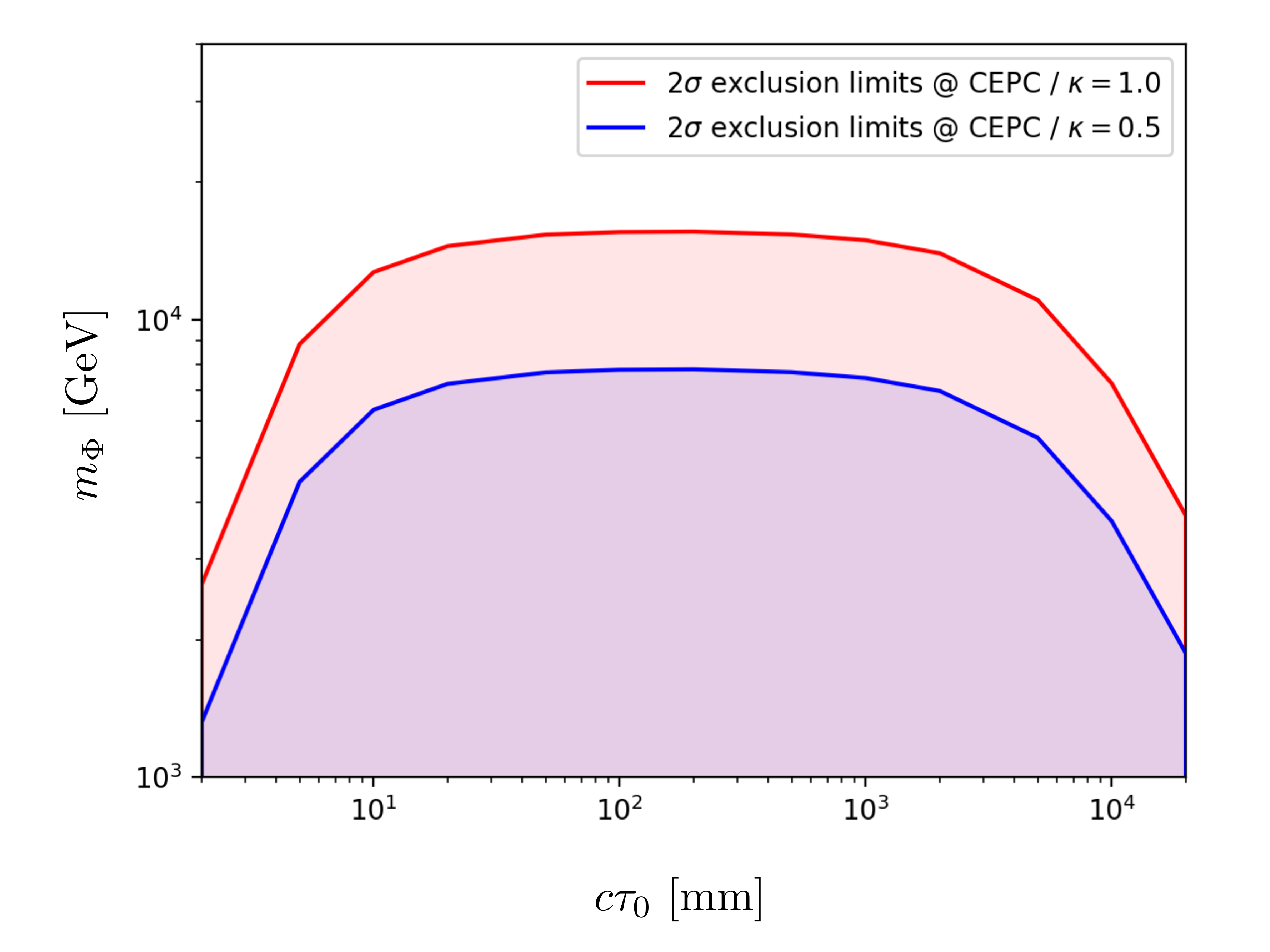}
}
\caption{Left: Illustration plot of the signal process at CEPC. Detector size is denoted by two circles. Black dotted lines and red solid lines are dark pions and muons, respectively.
Right: 2 $\sigma$ exclusion limits on mediator mass $m_{\Phi}$ as functions of the dark pion proper decay length, with coupling $\kappa$ fixed to 0.5 and 1.0 respectively.}
\label{CEPC2}
\end{figure}

In Fig.~\ref{CEPC1} (right) we present $\sigma(e^{+}e^{-}\to \bar{q'}q')$ as functions of $m_{\Phi}$ with $\kappa$ fixed to $0.5$ and $1.0$.
Thanks to the expected high luminosity, CEPC can produce plenty of $q'$ pair even with TeV scale mediator. 
Dark quark $q'$ finally evolve to a jet-like object, which is composed by lots of displaced muons, see Fig.~\ref{CEPC2} (left) for illustration.  
Detection of displaced signals is closely related to the physical size of detector. 
For CEPC, the inner radius of inner tracker (IT) is 16 mm, and the outer radius of muon spectrometer (MS) is 6.08 m. 
Based on these design information, here we propose a cut-flow to search for displaced lepton jets:
\begin{itemize}
\item Dark pion $\pi'$ decay at a place away from primary vertex, and thus generate a displaced vertex (DV). 
We denote the transverse distance from a DV to the primary vertex as $L_{xy}$.   
Muon coming from dark pion decay can be traced back to a DV. 
If the transverse distance of this mother DV is within the range $16 \text{ mm} < L_{xy} < 6.08 \text{ m}$,
then this muon will be tagged as a displaced muon. 
\item We use all the displaced muons, with $p_{\text{T}} > 1$ GeV and $|\eta|<3.0$, as input of jet clustering. We use anti-kt algorithm with jet radius $R=0.4$ to do jet clustering.  
If there are 6, or more than 6, displaced muons inside a jet, then this jet will be tagged as a displaced muon jet (DMJ). 
\item We require the number of DMJs to be greater than 2.
\end{itemize}

Unlike the LHC DMJ analysis, which is designed for FRVZ model and only require at least two muons inside a DMJ, our DMJ criteria require more displaced muons inside it. 
Thus our DMJ can not be faked by SM background processes. 
Furthermore, all the displaced muons can be paired by the DV which they belong to. 
So the reconstructed momentum of each displaced muon pair should point to the primary vertex. 
There are at least 6 this kind of displaced muon pair is our signal event. 
These amount of  muon pairs with their momentums point to the primary vertex, can hardly be faked by cosmic rays. 
So the search of our DMJ signal at CEPC can be treated as background free~\cite{Liang:2018mst,discussion}. 

If there is no background events, then the 2$\sigma$ exclusion limit corresponds 3 expected signal events. This is called ``rule of three'' is statistics.  
In Fig.~\ref{CEPC1} (right) we present the 2$\sigma$ exclusion limits on $c\tau_0 - m_{\Phi}$ plane with $\kappa$ fixed to $0.5$ and $1.0$.
In this exclusion plot we choose central energy $\sqrt{s}=240$ GeV and integrated luminosity 5.6 ab$^{-1}$ for CEPC setting. 
It can be seen that $m_{\Phi}$ can be excluded up to about 10 TeV scale at future CEPC, provided the proper lifetime of dark pion is in the range of 10 mm to 10 m, which just correspond to the physical size of CEPC detector.


Before we finish this section, we give a brief discussion on displaced electron jet signal.  
If we reduce the mass of $\pi'$ to less than 200 MeV, then the main decay channel of $\pi'$ will be $\pi'\to \bar{e}e$, and thus the collider signature for our model changes to displaced electron jets.  
At LHC, the displaced electron jets signal suffer from multi-jets contamination, and ATLAS observed 239 displaced electron jet pair events that mainly comes from multi-jets process~\cite{ATLAS:2016jza}. 
So the detection of our model at LHC will become more difficult if the signal is displaced electron jets. 
However, CEPC is much more cleaner than LHC, and we can expect that the contamination from multi-jets process can be well controlled at CEPC. 
Thus displaced electron jets will be a more promising signal for the detection of our model at CEPC. We will study this case in a future work.

\section{Conclusion}\label{sec:conclusion}

In this work we propose a composite asymmetric dark matter model. 
The dark sector in this model talks to the SM sector through a scalar mediator, which couples to a SM lepton and a dark quark.
This model can successfully explain the observed baryon asymmetry of the universe and dark matter relic density.
Detection of this model is briefly discussed. 
We find that current dark matter direct detection, whether via DM-electron scattering or DM-nucleon scattering, could hardly constrain this model. 
Furthermore, due to the small production cross section and irreducible background, current LHC is also incapable to detect this model. 
Finally, we find it is promising to detect this model by ``displaced lepton jets'' signal at future lepton colliders. 
Using CEPC as a representative lepton collider, we find that CEPC could exclude the mass of mediator up to 10 TeV scale, provided the proper lifetime of dark pion varies from 10 mm to 10 m.

\section*{Acknowledgements}
M.Z. thanks Junmou Chen, Zhen Liu, Michael J. Ramsey-Musolf, Manqi Ruan, and Fanrong Xu for useful discussions. 
M.Z. appreciates Fa Peng Huang for careful reading of this manuscript. 
This work was supported by the National Natural Science Foundation of China (NNSFC) under grant No. 11947118.

\vspace{0.5cm}

\bibliographystyle{apsrev4-1}

\end{document}